# Conjoined Charge Density Waves in the Kagome Superconductor $CsV_3Sb_5$

Haoxiang Li[1], G. Fabbris[2], A. H. Said[2], Y. Y. Pai[1], Q. W. Yin[3], C. S. Gong[3], Z. J. Tu[3], H. C. Lei[3], J.P. Sun[4,5], J.-G. Cheng[4,5], Ziqiang Wang[6], Binghai Yan[7], R. Thomale[8], H. N. Lee[1], and H. Miao[1,*]

[1]*Material Science and Technology Division, Oak Ridge National Laboratory, Oak Ridge, Tennessee 37831, USA*
[2]*Advanced Photon Source, Argonne National Laboratory, Argonne, Illinois 60439, USA*
[3]*Department of Physics and Beijing Key Laboratory of Opto-Electronic Functional Materials and Micro-devices, Renmin University of China, Beijing, China*
[4]*Beijing National Laboratory for Condensed Matter Physics and Institute of Physics, Chinese Academy of Sciences, Beijing 100190, China*
[5]*School of Physical Sciences, University of Chinese Academy of Sciences, Beijing 100190, China*
[6]*Department of Physics, Boston College, Chestnut Hill, Massachusetts 02467, USA*
[7]*Department of Condensed Matter Physics, Weizmann Institute of Science, Rehovot 7610001, Israel*
[8]*Institute for Theoretical Physics, University of Würzburg, Am Hubland, D-97074 Würzburg, Germany*

**The intricate interplay between novel lattice geometry and spontaneous symmetry-breaking states is at the forefront of contemporary research on quantum materials. Recently, the observation of unconventional charge and pairing density waves in a kagome metal $CsV_3Sb_5$ brings out a new showcase for intertwined orders. While electronic instabilities in $CsV_3Sb_5$ are widely believed to originate from the V 3$d$-electrons residing on the 2-dimensional kagome sublattice, the pivotal role of Sb 5$p$-electrons for 3-dimensional orders is yet to be understood. Here, using resonant tender x-ray scattering and high-pressure X-ray scattering, we report a rare realization of conjoined charge density waves (CDW) in $CsV_3Sb_5$. At ambient pressure, we discover a resonant enhancement at Sb $L_1$-edge ($2s\rightarrow 5p$) at the 2×2×2 CDW wavevectors. The resonance, however, is absent at the 2×2 CDW wavevectors. Applying hydrostatic pressure, we find the CDW transition temperatures to separate, where the 2×2×2 CDW emerges 4 K above the 2×2 CDW at 1GPa. Our results establish the coexistence of the 2×2 CDW and the 5$p$-electron assisted 2×2×2 CDW in $CsV_3Sb_5$. The evolution of the conjoined CDWs under pressure suggests the joint importance of electronic and phononic fluctuations for the double dome superconductivity.**

Charge density waves (CDWs), a translational symmetry-breaking electronic fluid state of matter, are in the spotlight to unravel intertwined quantum materials. This includes cuprate high-$T_C$ superconductors as well as Moiré superlattices[1-4] and, most recently, non-magnetic kagome metals where the CDW emerges as a leading electronic instability near van Hove filling[5-12]. Intriguingly, due to the sublattice interference on the geometrically frustrated triangle network, CDWs are predicted to potentially feature finite angular momentum or time-reversal symmetry breaking[7-12]. An exotic CDW in combination with superconductivity has been discovered in the kagome superconductor family $A$V$_3$Sb$_5$ ($A$=K, Rb, Cs) (Fig. 1a)[13-25]. At ambient pressure, a CDW in $A$V$_3$Sb$_5$ sets in between 78 K to 103 K. Concomitant electronic nematicity and time-reversal symmetry breaking have been observed[15,16,22,24,25]. Below $T_{SC}$~3 K, superconductivity (SC) emerges and displays an unconventional pair-density modulation in real space[23].

While it is apparent that the CDW plays a key role in describing the exotic electronic phases in AV$_3$Sb$_5$, the microscopic origin of the CDW and its impact on superconductivity are still to be discovered. Like the conventional vs. unconventional paradigm for superconducting pairing, a CDW can be mediated by electronic interactions or phonons which is challenging to discriminate. This particularly applies to electronic kagome bands, where the 3$d$-van Hove singularities at the M point imply particle-hole fluctuations favoring a 2×2 CDW (Fig. 1b) [5-9]. In experimental studies, the CDW in kagome metals appears to be 3-dimensional (3D) with a 2×2×2 superstructure[17,18], indicating a non-trivial interlayer coupling and non-negligible lattice effects[26]. Still, a Landau theory analysis readily shows that the 2×2×2 CDW alone is incompatible with the first order phase transition (Fig. 1c,d), which could already point to a novel coexistence of 2×2 and 2×2×2 CDWs[11,12]. Turning to the superconducting phase, the intimate correlation between CDW and superconductivity extends to the temperature ($T$) $vs$ pressure ($P$) phase diagram[20-22]. As depicted in Fig. 1a, the CDW transition temperature, $T_{CDW}$, monotonically decreases and eventually vanishes at $P_{c2}$. Near an intermediate pressure $P_{c1}$~0.7 GPa, the curvature of the resistivity at $T_{CDW}$ changes sign. Interestingly, superconductivity also shows two turning points at $P_{c1}$ and $P_{c2}$ and forms a double dome structure. Using resonant tender x-ray scattering and high-pressure X-ray scattering, we demonstrate that the 2×2 CDW and a 5$p$-electron assisted 2×2×2 CDW are coexisting and intertwined in CsV$_3$Sb$_5$. The conjoined CDWs we observe transcend the phenomenology that could be derived from a sole V kagome sublattice near van Hove filling, and hence provides key information to understand the enlarged complexity of novel symmetry-breaking phases in CsV$_3$Sb$_5$.

We start by proving the different electronic origin of the 2×2 and 2×2×2 CDWs using resonant elastic x-ray scattering (REXS). As depicted in Fig. 2a, by tuning the photon energy to atomic absorption edges, orbital-resolved valence electrons that are involved in the CDW will be enhanced[3]. In the cuprate high-$T_c$ superconductors, CDW intensity resonates at both O $K$-edge ($1s \rightarrow 2p$)[27] and Cu $L_3$-edge ($2p \rightarrow 3d$)[28,29], reflecting the $d$-$p$ hybridized wave function. Here we choose the Sb $L_1$ edge ($h\nu$=4.7 keV) to enhance the 5$p$-electrons near the Fermi level (see supplementary materials for REXS at other edges). Comparing with the V $L_3$ edge ($h\nu$=0.512 keV) that has constrained scattering in reciprocal space (see supplementary materials), the higher photon energy of the Sb $L_1$ edge allows us to reach both $\mathbf{Q}_{CDW}^{2\times2}$ with integer $L$ component and $\mathbf{Q}_{CDW}^{2\times2\times2}$ with half-integer $L$ component[18]. Moreover, since Sb 5$p_z$-electrons form a large Fermi surface and directly contribute to the coupling between adjacent kagome layers, the Sb $L_1$-edge resonance can potentially be used to distinguish the 2×2 CDW and the interlayer coupled 2×2×2 CDW. Our main observations at the Sb $L_1$ edge are summarized in Fig. 2c-f. All REXS data were collected in the reflection geometry at $T$ = 10 K (see Supplementary Materials). The Sb $L_1$ edge is identified at $h\nu$=4.7 keV in the fluorescence scan shown in Fig. 2c. Figure 2d and 2e reveal energy scans at $\mathbf{Q}_{CDW}^{2\times2} = (\pm 0.5, 0, 2)$ and $\mathbf{Q}_{CDW}^{2\times2\times2} = (\pm 0.5, 0, 2.5)$ (see more $\mathbf{Q}$ points in the supplementary materials). Remarkably, we find a strong dip in scans at $Q_{CDW}^{2\times2}$, which is in stark contrast to scans at $\mathbf{Q}_{CDW}^{2\times2\times2}$ that show large resonant enhancement slightly below the Sb $L_1$ edge. The dramatically different resonant response between $\mathbf{Q}_{CDW}^{2\times2}$ and $\mathbf{Q}_{CDW}^{2\times2\times2}$ demonstrates different electronic origins of the 2×2 and 2×2×2 CDWs in CsV$_3$Sb$_5$, where the Sb 5$p$ valence electrons are only involved in the formation of 2×2×2 CDW.

Having established the two conjoined CDWs with different charge contribution, we examine the CDW evolution under pressure. The experimental geometry of high-pressure diffraction is depicted in Fig. 3a. Figure 3b shows the integrated CDW intensity at $P$=0.5 GPa< $P_{c1}$. Consistent with the ambient pressure measurement, the CDW peaks at $\mathbf{Q}_{CDW}^{2\times2\times2}$ and $\mathbf{Q}_{CDW}^{2\times2}$ emerge at the same temperature around 77 K. Upon increasing pressure up to $P=1$ GPa$> P_{c1}$, however, the degeneracy of the conjoined CDWs is lifted with a 4 K gap between $T_{CDW}^{2\times2\times2}$ and $T_{CDW}^{2\times2}$ (Fig. 3c). Figure 3d and 3e compare the 2×2 and 2×2×2 CDW intensities at 57 K and 61 K, respectively. The absence of the 2×2 CDW peak at 61 K provides evidence for the conjoined nature of CDWs in CsV$_3$Sb$_5$. In particular, the 2×2×2 CDW peak that precedes the 2×2 peak at $P>P_{c1}$ cast doubt on any scenario of a simple stacking order with a π phase shift between 2×2 in-plane lattice distortions, as this should lead to a higher $T_{CDW}^{2\times2}$ compared to $T_{CDW}^{2\times2\times2}$. Instead, our observations indicate a 5$p$-electron assisted 3D CDW that could admix electronic and lattice fluctuations.

While the electronic density of states stemming from the 3$d$-van Hove singularity tends to favor a 2×2 CDW, there are different lattice distortions that could correspond to it. Experimental studies of $A$V$_3$Sb$_5$ found that the lattice distortion in the CDW phase is consistent with Star-of-David (SoD) or inverse-SoD patterns[16,30,31]. First principles calculations showed that the inverse-SoD is energetically favored at zero temperature but nearly degenerate with the SoD pattern at high temperature[32,33]. In the presence of Sb sublattices, both the kinetic energy ($p_z$-$p_z$ bonding) and Coulomb repulsion between 5$p$-electrons are sensitive to the lattice distortions. Figure 2b displays an anti-phase movement of Sb2 atoms that is derived from a SoD and inverse SoD stacking. This movement keeps the inter-layer Sb2-Sb2 distance and hence balances the kinetic and potential energies of 5$p$-electrons. Interestingly, already simplified model based on the 5$p$-electron stabilized SoD and inverse SoD stacking provides a possible lattice solution of the conjoined CDWs (see supplementary materials). The critical role of Sb 5$p$-electrons together with the precedent 2×2×2 CDW at $P>P_{c1}$ strongly indicates that the, at least partly, correlation-driven CDW state is superseded by lattice effects under higher pressure which favour the 2×2×2 pattern. Indeed, first principles calculations have shown that the hydrostatic pressure at 1GPa is sufficient to move the 3$d$-van Hove singularity away from $E_F$ and significantly suppress the charge susceptibility. With the results from REXS and high-pressure X-ray scattering, we demonstrate the coexistence of the 2×2 CDW and the 5$p$-electron assisted 2×2×2 CDW in CsV$_3$Sb$_5$. The observation of conjoined CDWs performatively goes beyond the minimal framework of kagome electronic bands near van Have filling, and thus provides critical information to resolve the persisting puzzle of CsV$_3$Sb$_5$.

## Methods

### Sample preparation and characterizations

Single crystals of $CsV_3Sb_5$ were grown from Cs ingot (purity 99.9 %), V powder (purity 99.9 %) and Sb grains (purity 99.999 %) using the self-flux method[34]. The mixture was put into an alumina crucible and sealed in a quartz ampoule under partial argon atmosphere. The sealed quartz ampoule was heated to 1273 K for 12 h and soaked there for 24 h. Then it was cooled down to 1173 K at 50 K/h and further to 923 K at a slowly rate. Finally, the ampoule was taken out from the furnace and decanted with a centrifuge to separate $CsV_3Sb_5$ single crystals from the flux. The obtained crystals have a typical size of $2 \times 2 \times 0.1$ mm$^3$. CDW transition is clearly observed in specific heat measurement as shown in the inset of Fig. 1d.

### Resonant elastic X-ray scattering

Resonant single crystal X-ray diffraction was performed at the 4-ID-D beamline of the Advanced Photon Source (APS), Argonne National Laboratory (ANL). The X-rays higher harmonics were suppressed using a Si mirror and by detuning the Si (111) monochromator. Diffraction was measured using a vertical scattering plane geometry and horizontally polarized (σ) X-rays. The incident intensity was monitored by a He filled ion chamber, while diffraction was collected using a Si-drift energy dispersive detector with approximately 200 eV energy resolution. The probed absorption edges are close in energy (4.1-5.5 keV); thus, the use of this detector is key to reject the fluorescence background. The sample temperature was controlled using a He closed cycle cryostat and oriented such that X-rays scattered from the (001) surface.

### High-pressure X-ray diffraction

High-pressure single crystal x-ray diffraction was also measured at the 4-ID-D beamline of the APS, ANL. High pressure was generated using a modified Merrill-Bassett-type diamond anvil cell[35] fitted with a pair of Boehler-Almax anvils of 800 μm culet diameter. A stainless-steel gasket was indented to 70 μm and a sample chamber of 400 μm diameter was laser cut. 4:1 methanol:ethanol was used as pressure media. Diffraction was measured in the transmission geometry in which X-rays penetrate through both diamond anvils and sample. The sample was cut into approximately $80 \times 80 \times 40$ μm$^3$ and oriented such that the c-axis is parallel to the X-ray direction when θ = 0 degree. Temperature was controlled using a He closed cycle cryostat. Pressure was calibrated as a function of temperature using the Au lattice constant[36] and controlled in-situ using a He gas membrane. X-ray energy of 20 keV was used to minimize the diamond anvil attenuation. The incident intensity was measured using a $N_2$ filled ion chamber, and diffraction was collected using a NaI scintillator.

### meV-resolution hard X-ray diffraction

High-precision X-ray scattering measurements shown in Fig. 1c,d was taken at 30-ID-C (HERIX), where the highly monochromatic X-ray beam of incident energy $E_i$ = 23.7 keV (l= 0.5226 Å) was focused on the sample with a beam cross section of 35×15 μm$^2$ (horizontal×vertical). The total energy resolution of the monochromatic X-ray beam and analyzer crystals was Δ$E$~1.5 meV (full width at half maximum). The measurements were performed in transmission geometry.

**Acknowledgement:** We thank Kun Jiang, Jiaqiang Yan and Raphael Fernandes for stimulating discussions. This research was sponsored by the U.S. Department of Energy, Office of Science, Basic Energy Sciences, Materials Sciences and Engineering Division (REXS and high-pressure x-ray diffraction). H.C.L. was supported by National Natural Science Foundation of China (Grant No. 11822412 and 11774423), Ministry of Science and Technology of China (Grant No. 2018YFE0202600 and 2016YFA0300504) and Beijing Natural Science Foundation (Grant No. Z200005) (Sample growth). J.G.C. and J.P.S. was supported by National Natural Science Foundation of China (Grants No. 12025408, No. 11904391) (High-pressure x-ray scattering). Z.Q.W is supported by the U.S. Department of Energy, Basic Energy Sciences Grant No. DE-FG02-99ER45747 (Theory). B.Y. acknowledges the financial support by the European Research Council (ERC Consolidator Grant, No. 815869) and the Israel Science Foundation (ISF No. 1251/19 and No. 2932/21) (Theory). This research uses resources (REXS and high-pressure x-ray scattering at beam line 4ID and meV-IXS experiment at beam line 30-ID) of the Advanced Photon Source, a U.S. DOE Office of Science User Facility operated for the DOE Office of Science by Argonne National Laboratory under Contract No. DE-AC02-06CH11357. Extraordinary facility operations are supported in part by the DOE Office of Science through the National Virtual Biotechnology Laboratory, a consortium of DOE national laboratories focused on the response to COVID-19, with funding provided by the Coronavirus CARES Act.

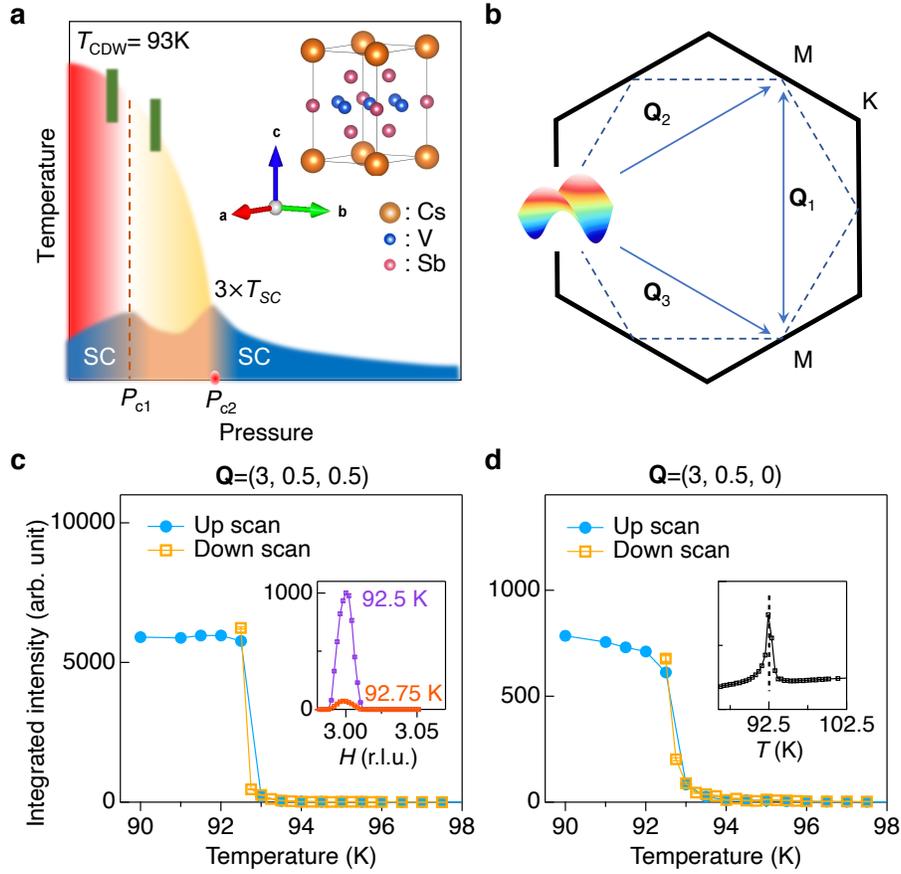

**Figure 1: Charge density wave in $CsV_3Sb_5$. a**, $T$-$P$ phase diagram and crystal structure of $CsV_3Sb_5$ (space group P6=mmm, no. 191). The phase diagram is seperated by two characteristic pressures, $P_{c1}$ and $P_{c2}$. The curvature of resistivity at $T_{CDW}$ changes sign at $P_{c1}$. At the same pressure, superconductivity reaches its first peak. The green markers indicate pressures where the high-pressure X-ray diffraction measurement were performed (Fig. 3). **b**, A schmatic of van Hove singularities at the M point of the hexagonal Brillouin zone. Theoretically, the van Hove filling induce Fermi surface instabilities with three nesting wavevector $Q_{1,2,3}$. **c-d**, Integrated CDW intensity vs temperature measured at $Q_{CDW}$=(3, 0.5 0.5) (panel **c**) and (3, 0.5, 0) (panel **d**) using meV-resolution elastic X-ray scattering. A sharp jump of the order parameter at the CDW transition indicate a weak first order phase transition and are consistent with nuclear magnetic resonance measurement[37,38]. The intensity jump is much stronger in the out-of-plane CDW peak at (3, 0.5, 0.5) compared to the in-plane one at (3.5, 0.5, 0). The inset of panel **c** shows the CDW peak intensity measured at 92.5 K and 92.75 K. The specific heat data shown in the inset of panel **d** reveal a sharp transition at 92.5 K, consistent with the diffraction data. The error bars in panels **c,d** represent 1 standard deviation assuming Poisson counting statistics.

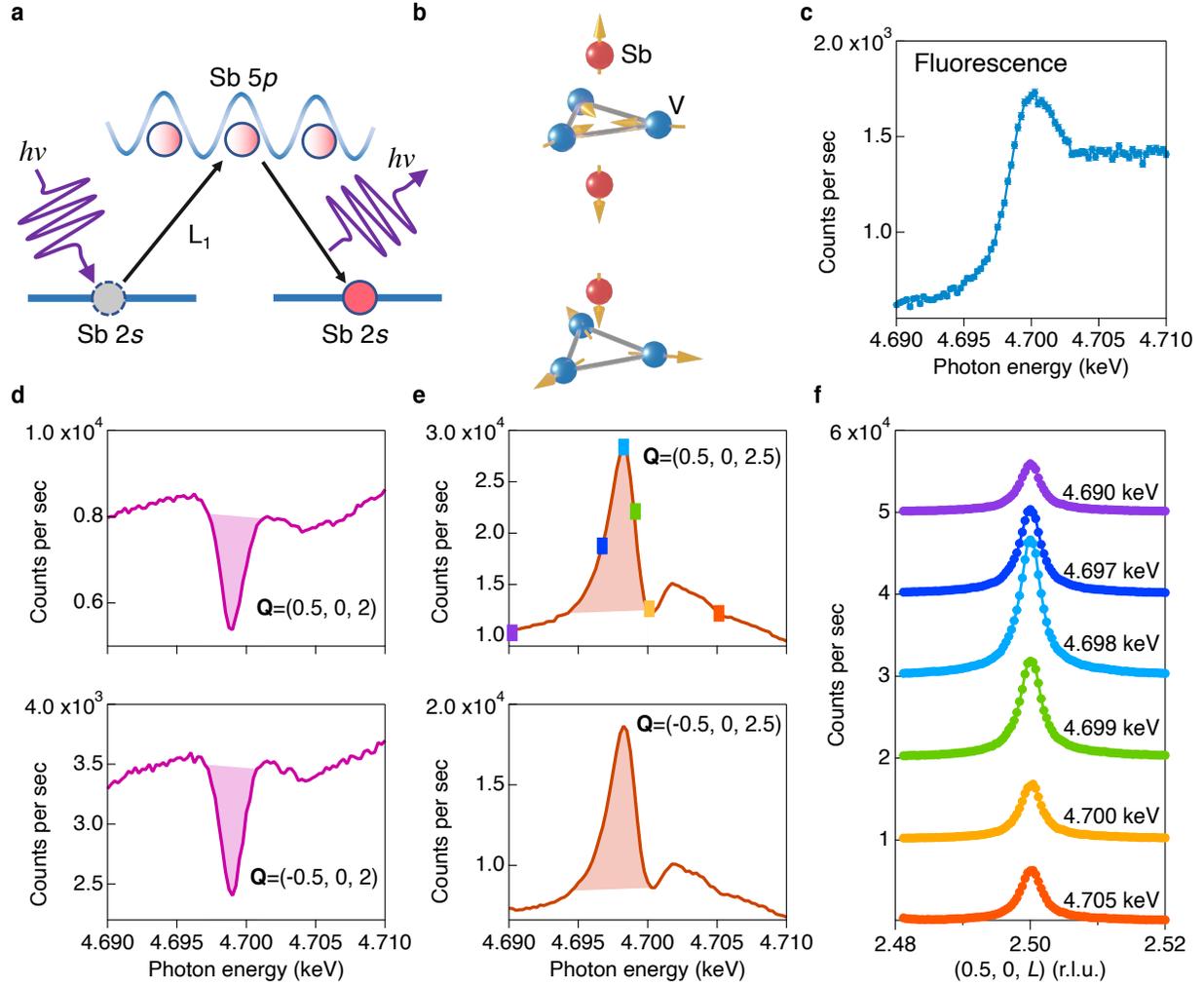

**Figure 2: Sb $L_1$-edge REXS reveals conjoined CDWs in CsV$_3$Sb$_5$. a,** Schematic of the REXS process. The Sb 2s to 5p transition is allowed by the dipole selection rule. **b**, Anti-phase movement of the Sb2 atoms that is derived from a SoD and inverse SoD stacking between adjacent kagome layers. **c**, X-ray fluorescence near the Sb $L_1$-edge (4.7 keV). **d-e,** Photon energy scans at the $\mathbf{Q}_{CDW}^{2\times2} = (\pm 0.5, 0, 2)$ and $\mathbf{Q}_{CDW}^{2\times2\times2} = (\pm 0.5, 0, 2.5)$ taken at $T$= 10 K. Resonant peaks in the energy scan of the 2×2×2 CDW demonstrate a Sb 5p-electron assisted CDW. This resonance can be directly observed in energy-dependent $L$-scans shown in panel **f**. The error bars in panel c represent 1 standard deviation assuming Poisson counting statistics.

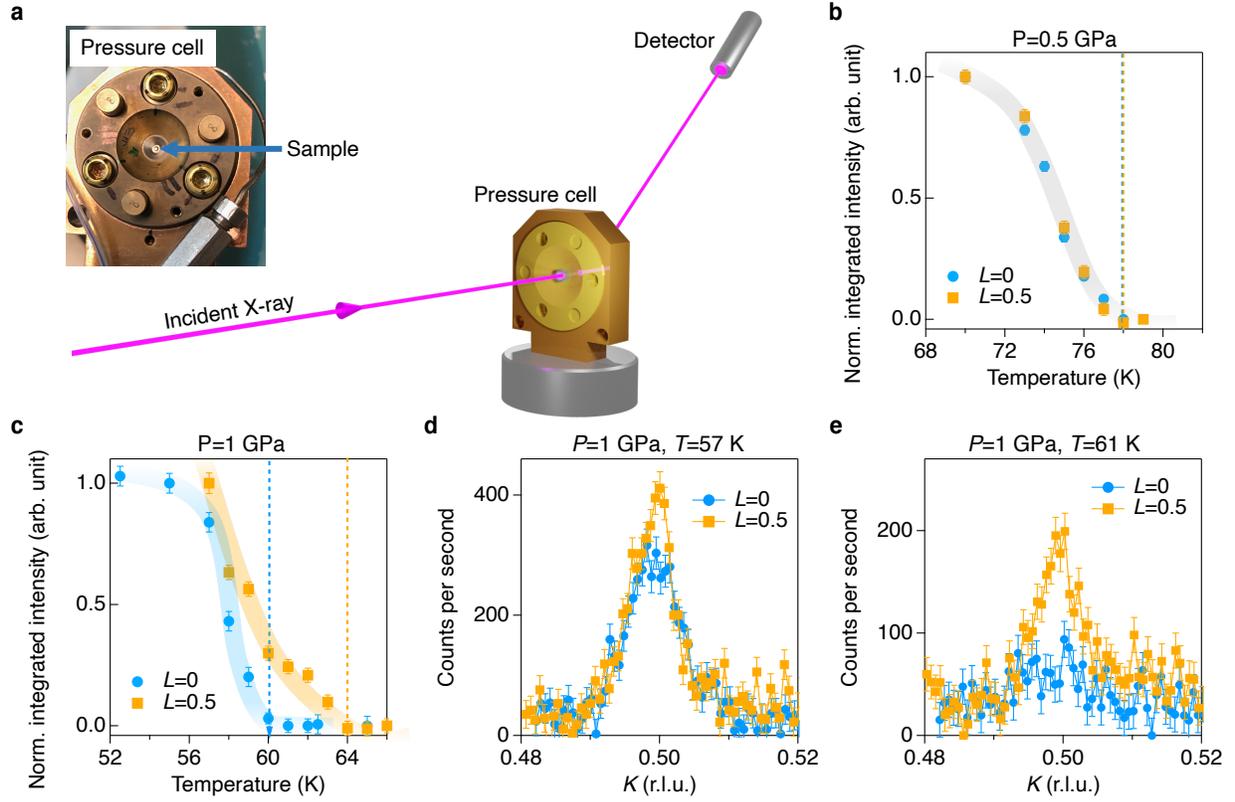

**Figure 3: Evolution of conjoined CDWs under pressure**. **a**, The experimental geometry of high-pressure X-ray diffraction. **b-c**, Normalized integrated CDW intensity vs temperature taken at $\mathbf{Q}_{CDW}^{2\times2\times2}=(3, 0.5, 0.5)$ and $\mathbf{Q}_{CDW}^{2\times2}=(3, 0.5, 0)$ at $P$=0.5 GPa and 1 GPa. Dashed lines mark the onset temperature of the 2×2×2 CDW (yellow) and 2×2 CDW (blue). The high-pressure X-ray scattering were taken at 20 keV in a transmission geometry. **d-e**, Direct comparison of the CDW peak intensity of $\mathbf{Q}_{CDW}^{2\times2\times2}=(3, 0.5, 0.5)$ and $\mathbf{Q}_{CDW}^{2\times2}=(3, 0.5, 0)$ under 1 GPa. The measurements are taken at $T$=57 K and 61 K. which is below and above the onset temperature ($T$=60 K) of the 2×2 CDW ($L$=0) under 1 GPa (blue dashed line in panel **c**). The error bars in panel **b-e** represent 1 standard deviation assuming Poisson counting statistics.